\documentstyle[12pt,epsf,epsfig]{article}
\def\beq{\begin{equation}}
\def\eeq{\end{equation}}
\def\bea{\begin{eqnarray}}
\def\eea{\end{eqnarray}}
\def\bq{\begin{quote}}
\def\eq{\end{quote}}

\def\gappeq{\mathrel{\rlap {\raise.5ex\hbox{$>$}}
{\lower.5ex\hbox{$\sim$}}}}

\def\lappeq{\mathrel{\rlap{\raise.5ex\hbox{$<$}}
{\lower.5ex\hbox{$\sim$}}}}

\begin{document}
\begin{center}

{\Large \bf Problems and Prospects in Spin Physics\\}
\vspace{5mm}
{John Ellis}\\
\vspace{5mm} 
{\small\it Theory Division, CERN, CH-1211 Geneva 23,
Switzerland}\\
\vspace{5mm}
ABSTRACT
\end{center}
\begin{center}
\begin{minipage}{130mm}\footnotesize{
This talk reviews some of the hot topics in spin physics and related
subjects, including perturbative QCD predictions for polarized parton
distributions and their possible behaviours at small $x$, the Bjorken and
singlet sum rules and the treatment of higher orders in perturbative QCD,
different interpretations of the EMC spin effect including chiral solitons
and the axial $U(1)$ anomaly, other experimental indications for the
presence of strange quarks in the nucleon wave function, implications for
dark matter physics, and a few words about polarization as a tool in
electroweak physics.}
\end{minipage}
\end{center}

\noindent{\Large \bf 1~~ Introduction and Outline}
\\
\par
Most of the talks at this meeting are concerned with
spin phenomena in the strong interactions, and
this emphasis is reflected in my talk, though I do have some
words to say at the end about polarization in the electroweak
interactions. Within QCD, we have a firm basis for understanding
polarization effects in perturbative QCD, whereas it is a puzzle
at the non-perturbative level which is linked to many fundamental
issues in the theory. Among the non-perturbative phenomena that may be 
illuminated by polarization experiments are chiral symmetry breaking
and the axial $U(1)$ anomaly. A bridge towards 
these non-perturbative effects may be provided by studies of higher
orders in QCD perturbation theory. For example, renormalons may guide
us in the identification of higher-twist and condensation phenomena in QCD.
Spin physics is also linked to other interesting phenomena in particle
physics, such as the question whether the proton wave function contains
many strange quarks, which would have implications for the possible 
violations of the Okubo-Zweig-Iizuka 
(OZI) rule seen in recent experiments using the
LEAR ring at CERN. Spin physics is also relevant to astrophysics and
cosmology, since the couplings of dark matter particles, such as 
neutralinos and axions, to ordinary baryonic matter involves the same
axial-current matrix elements that are measured in 
polarized-lepton-nucleon scattering experiments.

\par
The theoretical interest of these experiments is mirrored by the intense
experimental activity at many accelerator centres: CERN, SLAC, DESY, BNL,
the Jefferson National Accelerator Facility formerly known as CEBAF, etc..
In this talk I will try to bring out the puzzles found by experimentalists
to tease their theoretical colleagues, as well as the questions raised by
theory that require further experimental elucidation. Spin physics is in a
very active and exciting phase!

\par
The outline of my talk is as follows: I first review our understanding of
polarized partons in perturbative QCD, and flag their behaviour at small
$x$ as a theoretical issue in NNLO perturbative QCD that could be clarified
by data at HERA with a polarized proton beam. Then I discuss the evaluation
of QCD sum rules for polarized structure functions, addressing the issues
of the $Q^2$-dependence of the polarization asymmetry, where deep-inelastic
$\nu$-nucleon collisions could cast some light, and the resummation of
higher orders in QCD perturbation theory, which may be tackled using Pad\'e
approximants. Next, I emphasize that the interpretation of the EMC spin 
effect is an open issue in non-perturbative QCD, with chiral solitons and
non-perturbative $U_A(1)$ dynamics as competing explanations that 
require experiments to distinguish them. Here HERMES, COMPASS,
polarized RHIC and polarized HERA may be able to contribute via 
determinations of the gluon polarization $\Delta G$. On the other 
hand, experiments 
on the ``violation" of the OZI rule and final-state measurements in 
deep-inelastic lepton scattering may cast some light on the possible
existence and polarization of strange quarks and antiquarks in the 
nucleon wave function. Finally, after advertizing the connection
between polarization experiments and searches for non-baryonic 
dark matter, I close with praise for the r\^ole played by spin
physics in the electroweak sector, including precision measurements
at the $Z^0$ peak at both LEP and the SLC that suggest a mass range
for the Higgs boson within reach of forthcoming experiments at LEP 2 or
at the LHC.\\

\noindent{\Large \bf  2 ~~Polarized Partons in Perturbative QCD}
\\
\par
Since I am the first speaker at this meeting, it seems
that I must introduce the two polarized structure functions
$G_{1,2}(\nu, Q^2)$ that form the basis [1] for many of the
discussions in this talk and during the meeting:

\begin{equation}
\frac{d^2\sigma^{\uparrow \downarrow}}{dQ^2 d\nu} - 
\frac{d^2\sigma^{\uparrow \uparrow}}{dQ^2 d\nu} = 
\frac{4\pi \alpha^2}{Q^2 E^2}
\left [ m_N (E+E' \cos \sigma) 
G_1 (\nu , Q^2) - Q^2 G_2 (\nu , Q^2) \right ]
\end{equation}
In the Bjorken scaling limit $Q^2 \rightarrow \infty,
x \equiv Q^2/2 m_N \nu$ fixed, according to the naive
parton model:

\begin{equation}
m^{2}_{N} \nu G_1 {\nu , Q^2} \rightarrow g_1 (x),\,
m_N \nu^2 G_2 (\nu , Q^2) \rightarrow g_2 (x)
\end{equation}
The dominant $g_1(x)$ structure function is related in
the naive parton model to polarized quark distributions:

\begin{eqnarray}
g_{1}^{p} (x) & = & \frac{1}{2} \sum_{q} e^{2}_{q}
\left [ q_{\uparrow} (x) - q_{\downarrow} (x) +
\bar{q}_{\uparrow} (x) - \bar{q}_{\downarrow}(x) \right ] \nonumber \\
 & = & \frac{1}{2} \sum_{q} e^{2}_{q}
\Delta q (x)
\end{eqnarray}
whereas the unpolarized structure function $F_2(x)$ is given by:

\begin{equation}
F_2(x) = \sum_{q} e^{2}_{q} x
\left [ q_{\uparrow} (x) + q_{\downarrow} (x) +
\bar{q}_{\uparrow} (x) + \bar{q}_{\downarrow}(x) \right ]
\end{equation}
Polarized lepton-nucleon scattering experiments actually measure
directly the polarization asymmetry $A_1$ related to the virtual-photon
absorption cross sections $\sigma_{1/2, 3/2}$:

\begin{equation}
 A_{1} \equiv \frac{\sigma_{\frac{1}{2}} - \sigma_{\frac{3}{2}}}
{\sigma_{\frac{1}{2}} + \sigma_{\frac{3}{2}}}
\rightarrow
\frac{\sum_{q}e^{2}_{q} \left [q_{\uparrow} (x) - q_{\downarrow} (x) + ...\right ]}
{\sum_{q}e^{2}_{q}\left [q_{\uparrow} (x) + q_{\downarrow} (x) + ...\right ]}
\end{equation}

I will not address here the interesting issues related to the 
transverse asymmetry $A_2$, but move straight on to discuss the 
perturbative evolution of the polarized structure functions.

As $Q^2$ increases, the standard GLAP perturbative evolution
equations [2] take the form

\begin{eqnarray}
g_{1} (x,t) &  = & \frac{1}{2} <e^2> \int^{1}_{x} \frac{dy}{y} \times \nonumber\\
& \times & \left [ C^{S}_{q} \left ( \frac{x}{y} , \alpha_{s}(t) \right ) 
\Delta \Sigma (y,t) + 
2N_{f} C_{g} \left ( \frac{x}{y} , \alpha_{s} (t) \right ) \Delta G (y,t) \right.
\nonumber \\
 & + & \left. C^{NS}_{q} \left ( \frac{x}{y} , \alpha_{s} (t) \right )
\Delta  q^{NS} (y,t) \right ]
\end{eqnarray}
where $t \equiv \hbox{ln}(Q^2/\Lambda^2)$, the $C_{i}(x/y, \alpha_s(t))$
are coefficient functions, $\Delta \Sigma (y, t)$ is the singlet
combination of the $\Delta q(y, t)$, and $\Delta q^{NS}(y, t)$ is a
non-singlet combination. The polarized parton distributions obey
coupled integro-differential equations of the form

\begin{eqnarray}
\frac{d}{dt} \left ( 
\begin{array}{ll} \Delta \Sigma (y,t) \\ \Delta G (y,t)
\end{array} \right ) &  = &
\frac{\alpha_{s}(t)}{2 \pi} \int^{1}_{x} \frac{dy}{y}
\left ( \begin{array}{lr}
P^{S}_{qq} & 2N_{f}P_{qq}\\
P_{gq} & P_{gg}
\end{array} \right ) \left (\begin{array}{l} \Delta \Sigma \\ \Delta G
\end {array} \right ) \\
\frac{d}{dt} \Delta q^{NS}(y,t) & = &
\frac{\alpha_{s}(t)}{2 \pi} \int^{1}_{x} \frac{dy}{y}
P^{NS}_{qq} \left ( \frac{x}{y} , \alpha_{s}(t) \right )
\Delta q^{NS} (y,t)
\end{eqnarray}
where the $P_{ij}(x/y, \alpha_s(t))$ are polarized splitting functions. At leading order
in $\alpha_s(t)$, $C_q = 1, C_g = 0, P_{ij} = {\cal O}(1)$ [3]. Recently, thanks
to heroic efforts by local industry in particular [4], we now know the NLO
corrections 
$\delta (C, P) = {\cal O}(\alpha_s/\pi)$. The exact way in which these NLO  corrections
are divided up between the polarized quark and gluon distributions is
renormalization-scheme dependent, as we shall review later. There is much discussion at
this meeting of NLO QCD fits fits to polarized structure function data: one
state-of-the-art fit [5] is  shown in Fig.~1.

\begin{figure}
\hglue4.5cm
\epsfig{figure=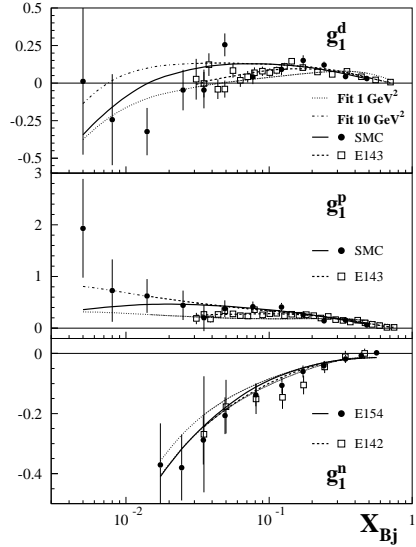,width=6cm}
\caption[]{
One state-of-the-art NLO perturbative QCD fit to polarized structure
function data, including lines showing the behaviours at two fixed values
of $Q^2$, and at the $x$-dependent values appropriate to different
experiments [5].}
\end{figure}

\par
One area where further guidance should be sought from perturbative QCD,
as well as information from experiment, is the low-$x$ region. Before
perturbative QCD came on the scene, the only guidance came from Regge
theory [6], which suggested that

\begin{equation}
g_{1}^{p,n} \simeq x^{-\alpha}
\end{equation}
where the intercept $\alpha$ of the relevant axial-vector meson
trajectory was guessed to lie in the range $0 \ge \alpha \ge -0.5$. Data
on the effective Pomeron intercept from HERA suggest that this might be
appropriate for $Q^2 < 1$ GeV$^2$, but that perturbative QCD plays an
essential r\^ole at higher $Q^2$. Standard resummation of the GLAP $Q^2$
logarithms $\Sigma_{n,m} \alpha_s(t)^n (\hbox{ln}Q^2)^m$ suggests [7] that

\begin{equation}
\left ( \ell n |\frac{1}{x}|\right )^{p} << g_{1}^{p,n} (x) << x^{-q}
\end{equation}
for any positive $p,q$. However, a BFKL-inspired resummation of $1/x$
logarithms $\Sigma_{n,m} \alpha_s(t_0)^n (\hbox{ln}1/x)^m$ at fixed $Q_0^2$
suggests [8] that

\begin{equation}
g_1(x,Q^2)^{NS} \sim x^{-0.4} \left ( \frac{Q^2}{\mu^2} \right )^{0.2} ,
g_1(x,Q^2)^{S} \sim x^{-1..0} \left ( \frac{Q^2}{\mu^2} \right )^{0.5}
\end{equation}
where we see that the singlet combination of structure functions
dominates the non-singlet at small $x$. BFKL behaviour is not required by the
unpolarized HERA data [9], and singlet dominance is not indicated by the SMC
polarized data [5] at the lowest available $x$. There is some discussion here 
of preliminary data from the SLAC E154 experiment, which can be fit by a 
power law: $g_1^n \sim x^{-0.8}$ for $0.02 < x < 0.1$, but these data are
at $Q^2$ too low, and $x$ too high, for equation (11) to be applicable.

\par
It has been shown that the resummation of the leading higher-order 
corrections is important for the singlet combination of polarized
structure functions, which is very sensitive to their treatment [10]. This is 
still an important source of systematic error, which could only be 
reduced significantly by tough theoretical calculations (NNLO, singular
terms at higher orders), and/or by measurements at polarized HERA [11]..\\

\noindent{\Large \bf  3 ~~Sum Rules}
\\
\par
Defining the integrals $\Gamma_1^{p,n} \equiv \int_0^1 g_1^{p,n}$, one 
has at the parton level the Bjorken [12] sum rule

\begin{equation}
\Gamma_{1}^{p} \, - \Gamma_{1}^{n} = \frac{1}{6} g_{A} = \Delta u - \Delta d
\end{equation}
whereas one can derive sum rules for the individual integrals

\begin{equation}
\Gamma_{1}^{p,n} = \left ( \pm \frac{1}{12} g_{A} + \frac{1}{36} a_{8} \right )
 + \frac{1}{9} \Delta \Sigma
\end{equation}
where $a_8 \equiv (\Delta u + \Delta d - 2 \Delta s)/\sqrt{3}$, only
if one assumes that $\Delta s = 0$ [13], implying for example that 

\begin{equation}
\Gamma_{1}^{p} \simeq 0.19
\end{equation}
The justification for this assumption was never very strong: to paraphrase Ref.
[13],  ``probably there are no strange quarks in the proton wave function, and if
there are, surely they are not polarized".

\par
The sum rules (12) and (13) acquire significant subasymptotic corrections
in perturbative QCD [14]:

\bea
\Gamma_{1}^{p} (Q^2) - \Gamma_{1}^{n} (Q^2)   &= & 
\frac{1}{6} g_{A} \left [ 1 - \frac{\alpha_{s}(Q^2)}{\pi} - 
3.5833 \left ( \frac{\alpha_{s}(Q^2)}{\pi} \right )^2 \right.\hfill \nonumber \\ 
&&\left.-   20.2153
 \left ( \frac{\alpha_{s} (Q^2)}{\pi} \right )^3 -
0(130) \left ( \frac{\alpha_{s}}{\pi} \right )^{4} + \ldots \right ]
\nonumber
\phantom{xxxxxxxxx}{(15a)} 
\eea
\bea
\Gamma^{p,n}_{1} (Q^2)  &=&  \left ( \pm \frac{1}{12} g_{A} + \frac{1}{36} a_{8} \right ) 
\left [ 1 - \frac{\alpha_{s}(Q^2)}{\pi} - 3.5833 
\left ( \frac{\alpha_{s}(Q^2)}{\pi} \right )^2 \right.\hfill\nonumber \\
&&\left. -  20.2153 \left ( \frac{\alpha_{s} (Q^2)}{\pi} \right )^3
- 0(130) \left ( \frac{\alpha_{s}}{\pi} \right )^{4} \right ] 
 + \frac{1}{9} \Delta \Sigma \hfill\nonumber \\
&&\left [ 1 - \frac{\alpha_{s}(Q^2)}{\pi} -
1.0959 \left ( \frac{\alpha_{s}(Q^2)}{\pi} \right )^2 -
0(4)\left ( \frac{\alpha_{s} (Q^2)}{\pi} \right )^3 + \ldots \right ]
\nonumber
\phantom{xxxx}{(15b)}
\end{eqnarray}
where the highest-order terms are estimates [15], 
leading, for example, to the prediction that, if $\Delta s = 0$, 
$\Gamma_1^p \simeq 0.17$ at a $Q^2$ typical of the EMC experiment [16].
\addtocounter{equation}{1}

\par
Sometimes there is still discussion of testing the Bjorken sum rule. This is
an absolutely fundamental prediction of QCD that has (in my opinion)
by now been tested successfully: for me, the issue is now to use
the subasymptotic corrections in the Bjorken sum rule
(15a) to determine $\alpha_s(Q^2)$ [17,18]. Using the subasymptotic 
corrections in (15b), one may check whether $\Delta s =0$, or measure
its value. A non-zero value is no ``crisis" for perturbative QCD, just
a surprise for overly naive models of non-perturbative QCD matrix elements.

\par
One of the issues to be confronted in testing these sum rules is that
the data are not all taken at the same $Q^2$, and some 
interpolation/extrapolation is necessary to evaluate $\Gamma_1^{p,n}(Q^2)$.
Experimentally, this is usually done by assuming that $A_1$ or $g_1/F_1$
is independent of $Q^2$, which seems empirically to be true [5] within the
errors for $Q^2 \gappeq 1$ GeV$^2$. Theoretically, there is no reason
why this should be true [7], although substantial deviations are not expected for
$x \gappeq 0.1$.  For larger $x$, the form of the expected 
scaling violations
is identical to NLO to those in $x F_3(\nu N)$, and it may be possible  to
use deep-inelastic $\nu N$ data to give hints how to interpolate/extrapolate
polarized $\ell N$ data where these are incomplete.

\par
The experimental determinations of the $\Gamma_1^{p,d}$ are compiled
in Fig.~2: things look grim for the singlet sum rules, but let us first
discuss quantitatively the Bjorken sum rule.

\begin{figure}
\mbox{\epsfig{figure=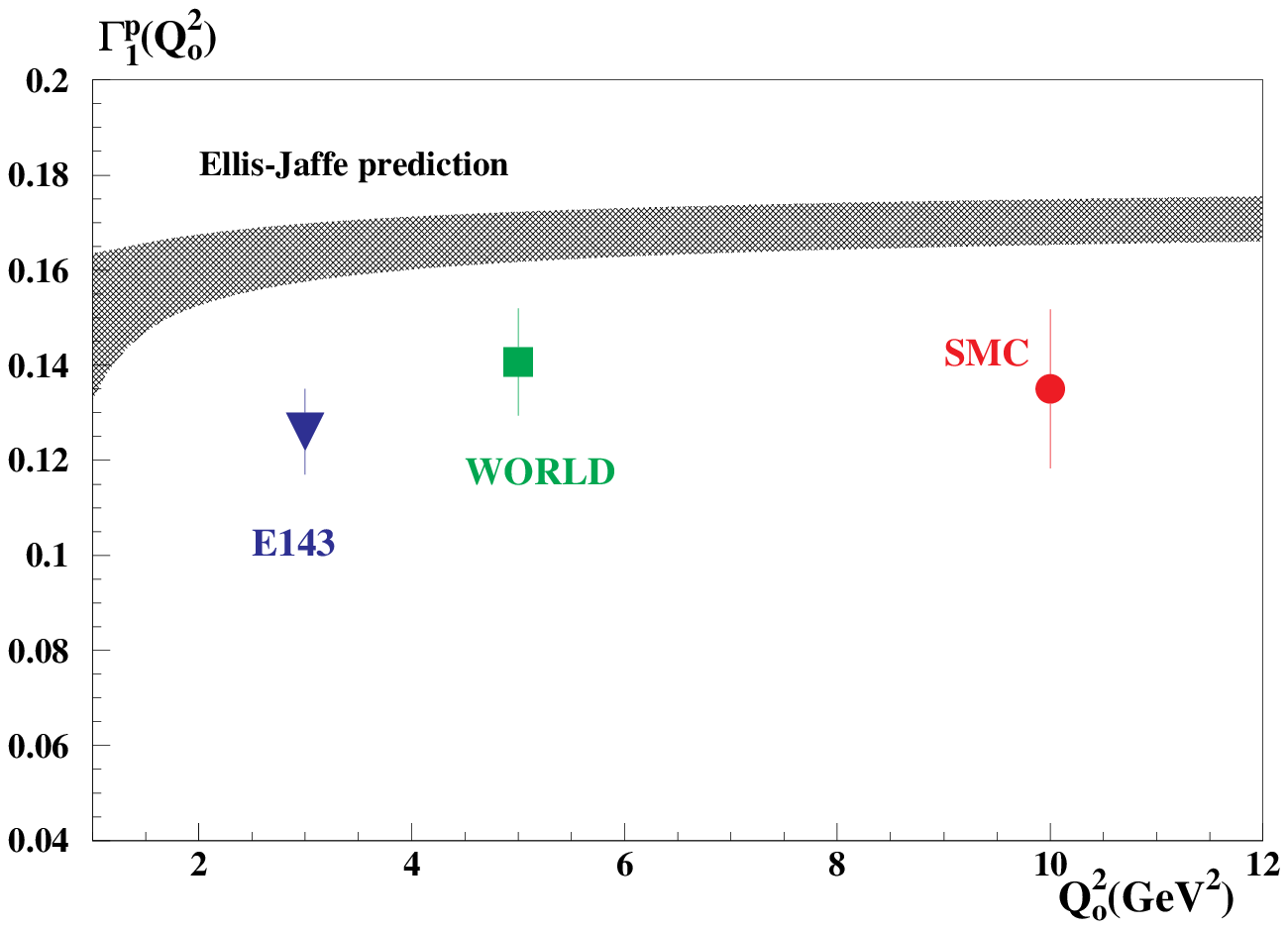,width=7.5cm}
\epsfig{figure=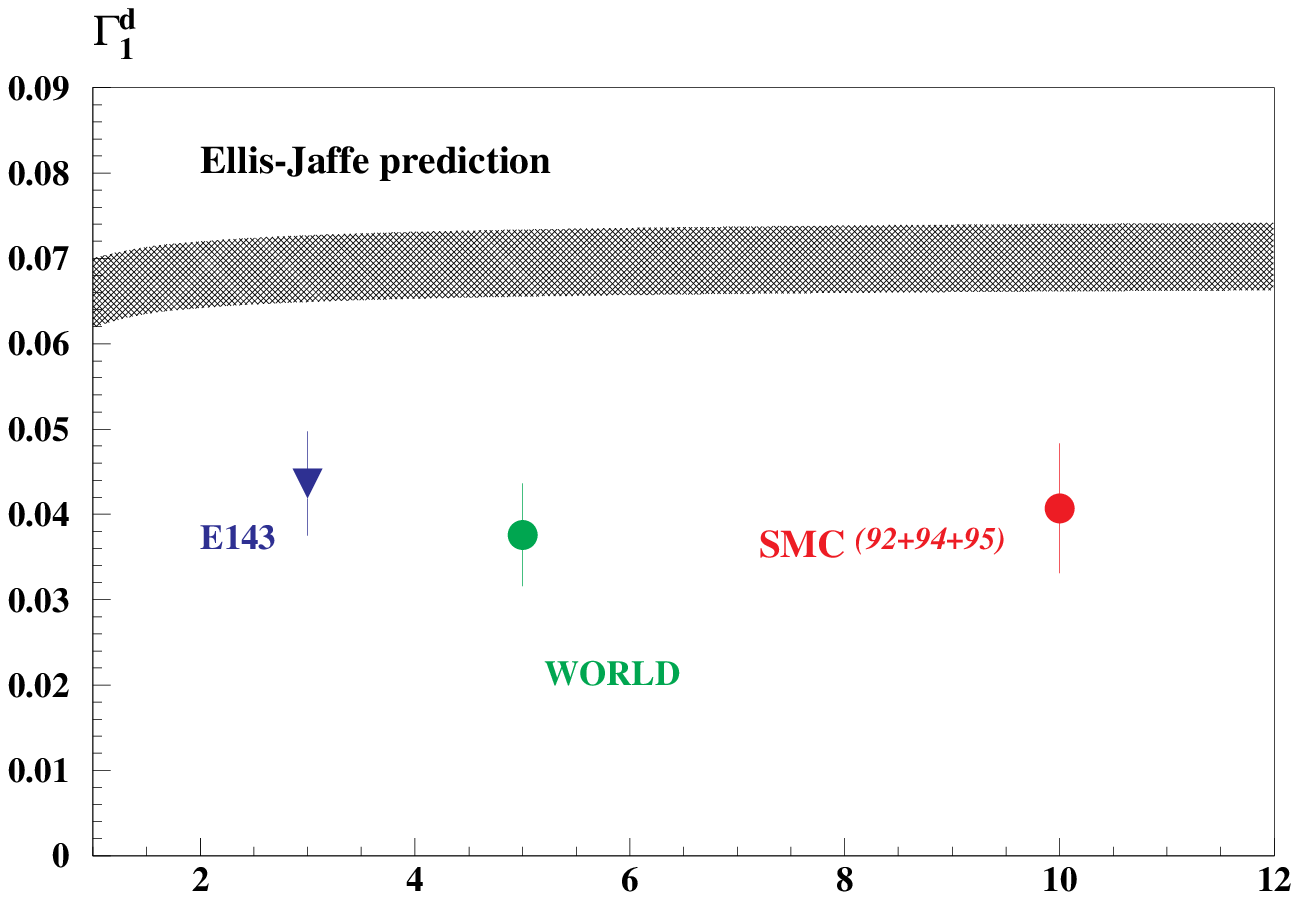,width=7.5cm}}
\caption[]{
A compilation [5] of measurements of the integrals
$\Gamma_1^{p,d}(Q^2)$, showing significant disagreement with the singlet
sum rule predictions [13].}
\end{figure}

\par
It is apparent from (15) that one of the issues that must be addressed
is the treatment of higher orders of QCD perturbation theory [18]. A 
generic QCD perturbation series is expected to be asymptotic:

\begin{eqnarray}
S(x) = \sum^{\infty}_{n=0} \, c_{n} x^{n} \, : \,
x \equiv \frac{\alpha_s}{\pi} , c_n \simeq n!K^n n^{\gamma}
\end{eqnarray}
corresponding to the presence of one or more renormalon singularities.
Confronted with an asymptotic series, one normally calculates it up to
the ``optimal" order $n_{opt} \equiv n: \Delta_n(x) \equiv |c_n x^n|$
is minimized, and then stops, quoting $\Delta_{n_{opt}}$ as an error
estimate. Looking at the series in (15), it seems that we have not yet
reached $n_{opt}$ for $\alpha_s / \pi \lappeq 1$. Is there some way of estimating
the next terms in the series, and of resumming the series so as to minimize the
estimated error?

\par
One approach is to use Pad\'e approximants:

\begin{eqnarray}
[N/M] = \frac{a_0 + a_1 x + \ldots a_N x^N}
{1 + b_1 x + \ldots b_M x^M} \, : \,
[N/M] = S + O(x^{N+M+1})
\end{eqnarray}
which are known to give good estimates in other applications [19], and
have been proven to converge if $\epsilon_n \equiv (c_n c_{n+2}/c^{2}_{n+1}) - 
1 \simeq {\cal O}(1/n)$, which is known to be the case for series 
dominated  by a finite number of renormalon poles [18]. We then predict
that

\begin{eqnarray}
\ell n|\delta_{[M/M]}|  \simeq  -M[1 + \ell n (2+a)]:
\delta_{[N/M]}  \equiv  \frac{c^{est}_{N+M+1} - c_{N+M+1}}
{c_{N+M+1}}
\end{eqnarray}
where $a$ is some number. This prediction is verified [1] for the
QCD vacuum polarization evaluated in the limit $N_f \rightarrow \infty$,
as seen in Fig.~3. Applied to the first three terms in (15a), the
$[1/2]$ and $[2/1]$ Pad\'e approximants yield [18] the estimate
$c_4 \simeq - 112 \pm 33$, to be quoted with the effective charge
estimate $c_4 \simeq - 130$.

\begin{figure}
\hglue3.5cm
\epsfig{figure=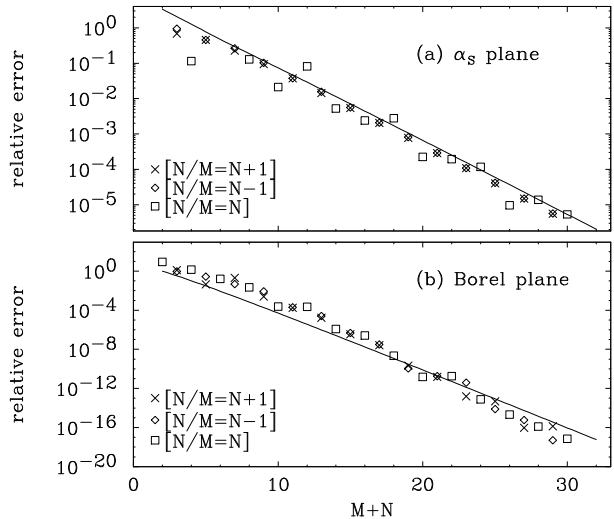,width=8cm}
\caption[]{
The errors in Pad\'e approximant predictions for the perturbative
expansion coefficients for the QCD vacuum polarization $D$ function in the
large-$N_f$ approximation, compared with theoretical error estimates
(18,20) [1].}
\end{figure}

\par
As seen in Fig.~3, the convergence of the Pad\'e
approximants is even better [18,19] when one makes a Borel transform of the
perturbative QCD series:

\begin{eqnarray}
\tilde{S} (y) \equiv \sum^{\infty}_{n=0} \, \tilde{c}_n y^n \, : \,
\tilde{c}_n \equiv \frac{c_{n+1}}{n!} \left ( \frac{4}{\beta_o} \right )^{n+1}
\, : \, \beta_0 \equiv \frac{(33 - 2N_f)}{3}
\end{eqnarray}
This is because ${\tilde \epsilon}_n \equiv ({\tilde c}_n {\tilde 
c}_{n+2}/{\tilde c}_{n+1}^2) - 1 \simeq 1/n^2$, implying that

\begin{eqnarray}
\ell n | \tilde{\delta}_{[M/M]}| \simeq -2M[1 + \ell n (2+a)]
\end{eqnarray}
as seen [1] in the second panel of Fig.~3. Indeed, if the series is
dominated by a finite set of renormalon poles, the Pad\'e
approximants become {\it exact}. In the case of the series (15a)
for the Bjorken sum rule, the $[2/1]$ Pad\'e approximant in the
Borel plane has
a pole at $y = 1.05$, to be compared with the expected renormalon
singularity at $y = 1$ [17].

\par
Further evidence for the reliability of Pad\'e approximants is
provided by studies of the renormalization scale and scheme
dependences of the Pad\'e resummation of the perturbative QCD series
for the Bjorken sum rule. We recall that the full ``sum" is a
measurable quantity that should be independent of renormalization scale:
$\partial/\partial \mu = 0$, and similarly of the scheme ambiguity
parametrized by $c_2$ at NLO: $\partial/\partial c_2 = 0$. Fig.~4
shows that the Pad\'e estimate of the sum is much less scale dependent
than partial sums of the QCD perturbation series [18], and Fig.~5 shows
that it also has very little renormalization-scheme dependence [20].

\begin{figure}
\hglue3.5cm
\epsfig{figure=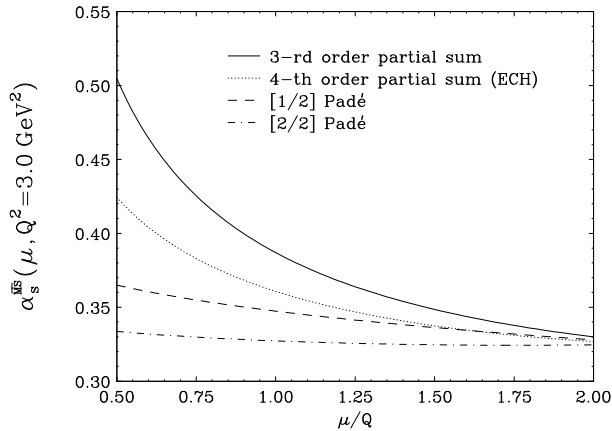,width=8cm}
\caption[]{
The renormalization scale dependence of the Bjorken sum rule is
greatly reduced by the use of Pad\'e approximants [18].}
\end{figure}

\begin{figure}
\hglue2.5cm
\epsfig{figure=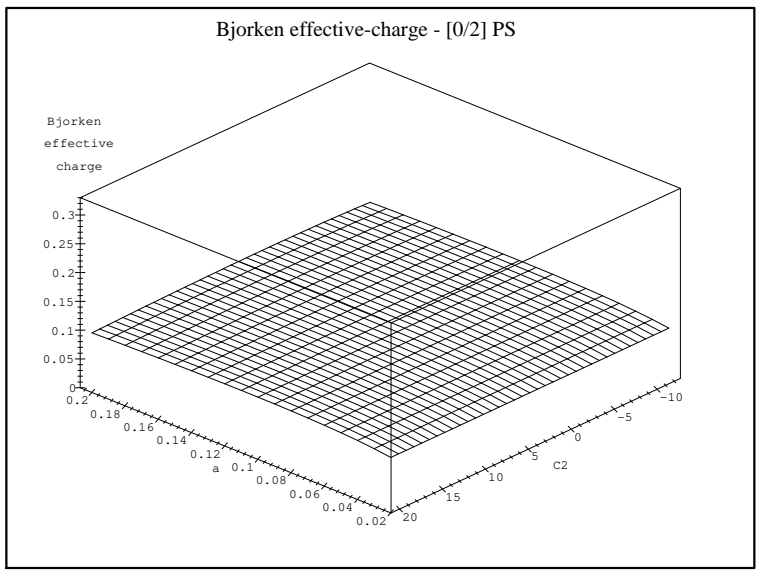,width=10cm}
\caption[]{
The renormalization scheme dependence of the Bjorken sum rule is also
greatly reduced by the use of Pad\'e approximants [20].}
\end{figure}

\par
We have used [20] the data available before this meeting in a numerical
analysis of the Bjorken sum rule, using Pad\'e approximants to sum
the QCD perturbation series. We took the data at face value, accepting 
the experimental estimates of their extrapolations to small $x$, and of
the associated systematic errors\footnote{It would be good to improve on the
treatment of the extrapolation  $x \rightarrow 0$ by combining the
proton and neutron data bin by bin in $x$, and using the best available
perturbative QCD formalism to extrapolate to $x = 0$.}. We do not
expect this extrapolation to be very badly behaved for non-singlet
combinations of structure functions, such as appear in the Bjorken sum
rule. Combining the experimental numbers, we find $\Gamma_1^p - 
\Gamma_1^n = 0.160 \pm 0.014$ at $Q^2 = 3$ GeV$^2$, corresponding to

\begin{eqnarray}
\alpha_s (M_z) = 0.117^{+0.004}_{-0.007} \pm 0.002
\end{eqnarray}
when using Pad\'e summation [20], where the first error is experimental,
and the second is a theoretical error associated with the residual
renormalization-scheme dependence in Fig.~5. This compares well with
the world average quoted at the Warsaw conference [21]: $\alpha_s(M_Z) = 0.118 
\pm 0.003$, and the consistency implies that the Bjorken sum rule is
verified at the $10\%$ level.

x

\par
We already saw in Fig.~2 that the singlet sum rules do not fare too well.
All the data are highly consistent if all the perturbative QCD 
corrections in (15) are taken into account [1]: this would not be apparent
for the neutron data if one did not include all the corrections. A global
fit to the data indicates that

\begin{eqnarray}
\Delta u & = & 0.81 \pm 0.01 \pm , \; \Delta d = -0.44 \pm 0.01 \pm , \nonumber \\
\Delta s & = & -0.10 \pm 0.01 \pm , \; \Delta \Sigma = 0.27 \pm 0.04 \pm
\end{eqnarray}
at $Q^2 = 3$ GeV$^2$. The second $\pm$ signs in (22) represent further
sources of error, including higher twists and the possible 
$Q^2$-dependence of $A_1$, but principally the low-$x$ extrapolation.
As already mentioned, this is very sensitive to the treatment of
higher-order perturbative QCD terms. Considerably more theoretical and
experimental work is necessary before the second errors in (22) can be
reduced to the level of the first ones. Nevertheless, it is clear that
$\Delta s \ne 0$ and that $\Delta \Sigma << 1$, in disagreement with
naive quark model estimates.\\

\noindent{\Large \bf  4 ~~Interpretations of the EMC Spin Effect}\\
\par 
{\bf Chiral Soliton (Skyrmion)}: CQCD has a large chiral symmetry : $q_{L(R)}
\rightarrow L(R) q_{L(R)}$, which has an $SU(N_f) \times SU(N_f) \times U(1)_B$
group structure that is broken  spontaneously by non-perturbative vacuum
expectation values
$<O|\bar{q}_{L}\bar{q}_{R} + h.c. |0>$ that couple the left $(L)$ and right
$(R)$ helicities [22]. We discuss later the fate of the axial $U(1)$ factor: it
is broken by an amount $O(1/N_c)$.  This spontaneous symmetry breakdown is
accompanied by the appearance of an octet of light pseudo-Goldstone bosons
$\pi^{\pm , 0}, K^{\pm , 0, \bar{0}}, \eta_{8}$, but not the singlet $\eta_0$
that disappears along with the axial $U(1)$ current: it decouples as $N_c
\rightarrow \infty$.  There is a simple effective Lagrangian [22]

\begin{eqnarray}
{\cal L}_{\pi} = \frac{f^{2}_{\pi}}{16} \, Tr (\partial_{\mu} U \partial^{\mu}
U^{\dagger}) +  \ldots : U \equiv \mbox{exp} \left [\frac{2i}{f_{\pi}} \underline{\pi}
\cdot \underline{\tau} \right ]
\end{eqnarray}
that describes their dynamics at energies $E<< \Lambda_{QCD}$ or distances $R
>> 1/\Lambda_{QCD}$ in the limit that $m_{u,d,s} << \Lambda_{QCD}$.  This Lagrangian
(22) has classical soliton (lump) solutions $U(\underline{x})$ [23] that are
labelled by the group $\pi_3(SU(N_f)) = {\bf Z}$, where the integers count the
baryon number $B$ that may be represented by

\begin{eqnarray}
B = \frac{1}{24\pi^2}\, \epsilon_{ijk} \, \int d^3 \underline{x} \, Tr
(\partial_i UU^{\dagger} \partial_j UU^{\dagger} \partial_k UU^{\dagger})
\end{eqnarray}
The lowest-lying state has $B = 1$ and may always be quantized as a fermion.

\par
Axial-current matrix elements in this model are given via PCAC by generalized
Goldberger-Treiman relations to the couplings  of the corresponding light mesons
$\pi^{\pm , 0}, K^{\pm , 0, \bar{0}}, \eta_8$.  There is no singlet axial-current
coupling because there is no singlet meson $\eta_0$ in the Lagrangian: the soliton is
a ``lump" of octet fields, and the $\eta_0$ decouples in the large-$N_c$ limit, in
which [24] 

\begin{eqnarray}
\Delta \Sigma = \Delta u + \Delta d + \Delta s = 0,
\end{eqnarray}
and there is also no way to obtain $\Delta G \neq 0$.

\par 
In the chiral soliton model, the nucleon spin arises [24] from the quantization
of the classical ``lump", and is interpreted as due to a coherent rotation of the
cloud of
$\pi , K$ and $\eta_8$ mesons : $L_z = \frac{1}{2}$. In the quark language, the
baryons are viewed as coherent states of a very large number of light, relativistic
quarks.  In the real world with $N_c = 3$ and $m_{u,d,s} \not=0$, we do not expect
(25) to be exact, but it might be accurate at the $O(1/N_c) \simeq 30 \%$ level, as
indicated by the experimental determination (22).

\par
Alternative approaches are based on the {\bf axial $U(1)$ anomaly}, which
contributes [25] to the perturbative evolution of the polarized parton
distributions at NLO:

\begin{eqnarray}
<p|\bar{q} \gamma_{\mu} \gamma_{5} q|p> = 2S_{\mu}
\left (\Delta q^{LO} - \frac{\alpha_s}{2\pi} \Delta G^{LO} \right )
\end{eqnarray}
As already mentioned, there is some freedom of renormalization-scheme choice at NLO,
and the $\overline{MS}$ option is simply to absorb the $\Delta G^{LO}$ correction
in (26) into the definition of $\Delta q$ at NLO: $\Delta q \equiv \Delta q^{LO}
- \frac{\alpha_s}{2\pi} \Delta G^{LO}$.  Alternatively, one may keep the $\Delta G^{LO}$
correction separate: $\Delta q, G^{LO} = \Delta q, G$, in which case 
$\Delta \Sigma = \Delta u + \Delta d + \Delta s - \frac{3\alpha_s}{2\pi} \Delta G$.
One can then hope to save the original assumption that $\Delta s = 0$ by
postulating that the apparent non-zero value of $\Delta s$ in (22) is in fact due to
the $\Delta G$ correction, and that $\Delta u + \Delta d + \Delta s$ is correctly
predicted by the naive non-relativistic quark model, with the small observed value of
$\Delta \Sigma$ also due to the $\Delta G$ correction.  This would require a rather large
positive value of $\Delta G$, which would need to be compensated by a correspondingly
large negative value of $L_z$.  Note that this axial-anomaly interpretation [25]
does not predict the measured values of $\Delta s$ and $\Delta G$ without an
extra dynamical assumption \footnote{For an alternative non-perturbative axial
$U(1)$ approach, see Ref. [26].}. 

\par
The first attempt to measure $\Delta G$ directly was in a Fermilab experiment [27]
looking for a particle-production asymmetry in polarized $pp$ collisions.  Their
measurements indicate that $\Delta G$ may even be negative \footnote{Some model
calculations [28] have also found $\Delta G < 0$.}, though they are not very conclusive,
because of the statistical and systematic limitations of the experiment.  There have
recently been several attempts [29] to extract $\Delta G$ indirectly from scaling
violations in polarized structure functions.  These hint that  $\Delta G > 0$, and are
mostly consistent with
$\Delta G = O(1)$.  However, the numerical value is still unknown:  uncertainties are still
large, and do not yet permit even an unambiguous measurement of the sign of $\Delta G$. 
This is the task entrusted to new experiments such as HERMES, COMPASS, polarized RHIC and
polarized HERA.\\

\noindent{\Large \bf  5 ~~``Breaking" the Okubo-Zweig-Iizuka Rule}
\\
\par
According to this rule [30], one should draw diagrams with connected quark lines, and
disconnected diagrams require gluon exchanges and should be suppressed.  The only
``breaking" of the rule for $\phi$ and $f'_2$ meson production is blamed on the small
$\bar{u}u + \bar{d}d$ components in their wave functions.  The rule works well for
mesons, but there is no good evidence that it works well for baryons. For example, if one
estimates the $\pi$-nucleon $\sigma$ term 
$\Sigma^{\pi N} = \frac{1}{2} (m_u + m_d) <p|(\bar{u}u + \bar{d}d)|p>$ using the
Gell-Mann-Okubo mass formula and the dynamical assumption that $<p|\bar{s}s|p> = 0$, one
finds $\Sigma^{\pi N} \simeq 25$ MeV, to be compared with the value $\Sigma^{\pi N} \simeq
45$ MeV inferred from experiment [31]. This corresponds to

\begin{eqnarray}
\frac{<p|\bar{s}s|p>}
{<p|(\bar{u}u + \bar{d}d +\bar{s}s)|p>} \simeq 0.2
\end{eqnarray}
Likewise, the polarized structure function data (22) indicate that 
$<p|\bar{s}\gamma_{\mu} \gamma_{5} s|p> \not= 0$, as do data [32] on elastic $\bar{\nu}
p
\rightarrow \bar{\nu}p$ scattering \footnote{Charm production in deep-inelastic $\nu N$
scattering also indicates the need for an $\bar{s}s$ component in the proton wave
function, though perhaps only at large $Q^2$.}.

\par
Confirming earlier suggestions [33], recent data from experiments at LEAR find large
apparent violations of the OZI rule in $\bar{p}N$ annihilation at
rest \footnote{Dispersion-relation analyses [34] of nucleon form factors are also
consistent with a large $\phi \bar{p}p$ coupling.}. The production ratios $\phi X/wX$
for $X =
\pi^{\pm ,0}, \pi \pi$ and $\gamma$ all exceed the OZI predictions, in some cases by as
much as a factor of 100! Moreover, there are indications that this OZI violation is
spin-dependent, since it is large in $S$-wave annihilations, but not in $P$-wave
annihilations, as seen in Fig. 6.

\begin{figure}
\hglue2.5cm
\epsfig{figure=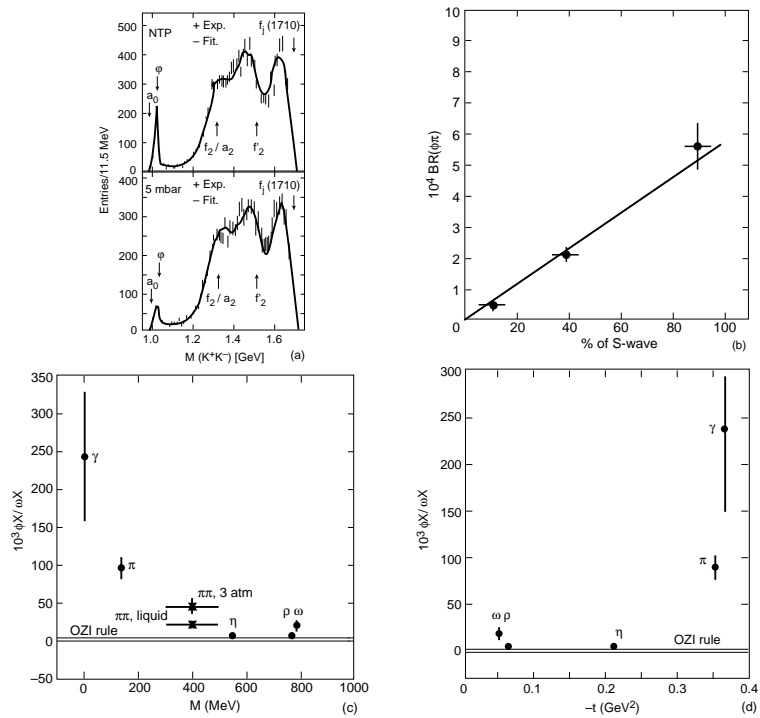,width=10cm}
\caption[]{
The $\phi$ production [35] in ${\bar p} N$ annihilation (a) depends on
the way in which the initial state is prepared, corresponding (b) to a
large enhancement of production in the $S$-wave annihilations favoured at
NTP. It also depends (c) on the final-state particle with which the $\phi$
is produced, and (d) on the momentum transfer. These are all features that
models [36] need to explain.
}\end{figure}

\par
One interpretation [33,36] of these data is that they are due to OZI evasion, rather
than breaking. If the proton wave function contains $\bar{s}s$ pairs, there are
additional connected quark-line diagrams that can be drawn, giving enhanced $\phi$
production. Such an $\bar{s}s$ component may also explain the backward peak seen in the
$\bar{p}p \rightarrow K^- K^+$ reaction.  Detailed models [36] agree qualitatively with
the pattern of OZI ``breaking" seen in the LEAR experiments. Other interpretations of
the apparent OZI ``breaking" include rescattering by intermediate $K$ and $K^*$ states,
but this is subject to systematic cancellations and does not seem to explain the
initial-state dependence that has been seen. There is also no trace of an exotic
resonance postulated in another interpretation of early data on OZI ``breaking".

\par
The proposition that the proton wave function contains $\bar{s}s$ pairs, and that they
are polarized, can be tested by looking for $\Lambda$ polarization in the target
fragmentation region of deep-inelastic $\nu ,e$ or $\mu N$ scattering [37].  The idea is
that the $\nu$ or polarized $e, \mu$ selects preferentially one parton polarization in
the proton wave functions, e.g., a $u^{\uparrow}$.  The proton remnant should remember
the polarization of the struck parton, and in particular one might expect negative $s$
polarization: $|p> \ni u^{\uparrow} (\bar{s}s)_{\downarrow} + \ldots$. Memory of this $s$
polarization may well be transferred with a dilution factor $D$ to any $\Lambda$ in the
current fragmentation: $s_{\downarrow} \rightarrow \Lambda_{\downarrow}$.  Such
$\Lambda$ polarization was indeed seen in the WA59 $\bar{\nu} N$ experiment [38]: for
$0.2 < x < 1$, $P_{\Lambda} = - 0.85 \pm 0.19$ to be compared with a postdiction
$P_{\Lambda} = -0.94 D$, indicating that $D = 0.9 \pm 0.2$.  This model can be used to
make predictions for polarized $e(\mu )N$ scattering: $P_{\Lambda} = 0.7 P_{\mu} D$, which
should be observable in the HERMES and COMPASS experiments.

\par
It would be interesting to know whether polarized-gluon models have anything to say
about OZI ``breaking" or $\Lambda$ polarization experiments.\\
\vfill\eject

\noindent{ \Large \bf  6 ~~Spin-off for Dark Matter Particles}
\\
\par
Polarization experiments are relevant to astrophysics and cosmology, as we now discuss.
One of the favoured candidates for cold dark matter is the lightest neutralino $\chi$,
which has spin-dependent couplings with nucleons that would be responsible for its
capture by the Sun (which could be detected by high-energy solar neutrinos produced by
$\chi \chi$ annihilation), and would contribute to elastic $\chi$ scattering off nuclei in
the laboratory (which could be used to detect dark-matter $\chi$ particles directly). The
spin-dependent matrix element $M$ contributing to $\sigma (\chi p \rightarrow \chi p)$ is
related to axial-current matrix elements [39].  In particular, if the $\chi$ particle is
approximately a photino

\begin{eqnarray}
M \simeq \sum_{q} e^{2}_{q} \Delta q
\end{eqnarray}
which is exactly the linear combination measured in polarized deep-inelastic $(e, \mu)p$
scattering experiments: have EMC et al. been measuring elastic $\chi p$ scattering?

\par
Another popular cold dark matter candidate is the axion, a very light pseudoscalar boson:

\begin{eqnarray}
m_a \propto m^{2}_{\pi} / f_a, \; g_{aff} \propto m_f / f_a, \; g_{a\gamma \gamma}
\propto 1 / f_a
\end{eqnarray}
where $f_a$ is a decay constant analogous to $f_{\pi}$, which was postulated to ensure CP
conservation in the strong interactions.  Accelerator experiments tell us that $f_a
\gappeq 1$ TeV, and coherent axion waves could contribute a significant fraction of the
mass density of the Universe if $f_a \sim 10^{11}$ to $10^{12}$ GeV. There are many
astrophysical bounds on $f_a$ which depend on the axion-nucleon couplings $C_{ap, an}$.
These are related by generalized Goldberger-Treiman relations to axial-current matrix
elements [40]:

\begin{eqnarray}
C_{ap} & = & 2(-2.76 \Delta u - 1.13 \Delta d + 0.89 \Delta s) - \cos 2\beta
 (\Delta u - \Delta d - \Delta s) \nonumber \\
C_{an} & = & 2(-2.76 \Delta d - 1.13 \Delta u + 0.89 \Delta s) - \cos 2\beta
(\Delta d - \Delta u - \Delta s)
\end{eqnarray}
where $\cos 2\beta$ is related to an unknown ratio of Higgs v.e.v.'s. Using (29), one can
determine the supernova axion emission rate, which is $\propto C^{2}_{an} + 0.8
(C_{an} + C_{ap})^{2} + C^{2}_{ap}$, as seen in Fig. 7. The determinations (21) of the
$\Delta q$ determine [17] the $C_{ap,an}$ with errors that are smaller than many other
astrophysical uncertainties, such as the equation of state inside a supernova core. 
Reconciling the axion emission rate with observations of neutrinos emitted by supernova
SN1987a is possible only if $f_a \gappeq 10^{10}$ GeV [40].  This is a second example of
the importance of polarization experiments for astrophysics and cosmology.\\

\begin{figure}
\hglue3.5cm
\epsfig{figure=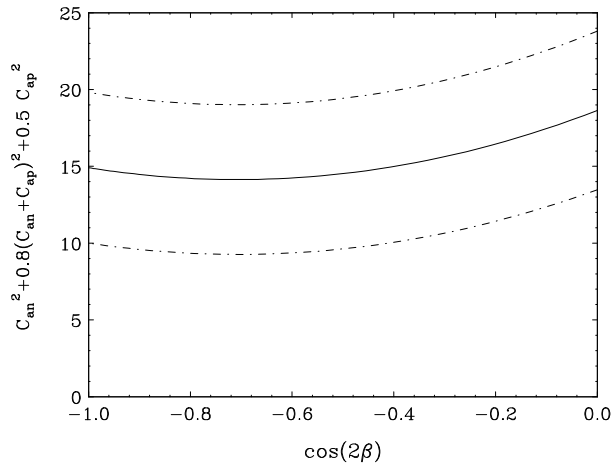,width=8cm}
\caption[]{
The axion emission rate from a supernova core is roughly proportional
to the scale on the vertical axis. Determinations of the $\Delta q$ now
constrain it with the uncertainty indicated by the dot-dashed curves [17],
which is considerably smaller than other uncertainties associated with
modelling the nuclear equation of state.
}\end{figure}

\noindent{\Large \bf  7 ~~Polarization in Electroweak Physics}
\\
\par
Although most of this meeting is concerned with spin phenomena in the strong
interactions, I cannot resist saluting briefly the importance of polarization in
electroweak physics. {\bf Transverse beam polarization} has been an invaluable tool
for calibrating the LEP beam energy and hence measuring accurately the $Z^{0}$ mass. One
of the most precise determinations of the electroweak mixing parameter $\sin^{2}
\theta_{W}$ is the left-right production asymmetry measured using {\bf longitudinal
beam polarization} at the SLC [41].  Other important measurements of 
$\sin^{2} \theta_{W}$ are made using {\bf $\tau$ polarization} measurements. 
Precision electroweak measurements enabled the top quark mass to be predicted on the
basis of radiative corrections. They are now able to predict [42] the Higgs boson mass
with a factor of 2 error:

\begin{eqnarray}
M_H = 145^{+ 164}_{- 77} ~~~~ GeV
\end{eqnarray}
as seen in Fig. 8. 
Perhaps future electroweak polarization measurements will enable us to refine further
this prediction.  If this prediction is confirmed by observation of the Higgs boson,
electroweak polarization will indeed have realized its rosy prospects!

\begin{figure}
\hglue2.5cm
\epsfig{figure=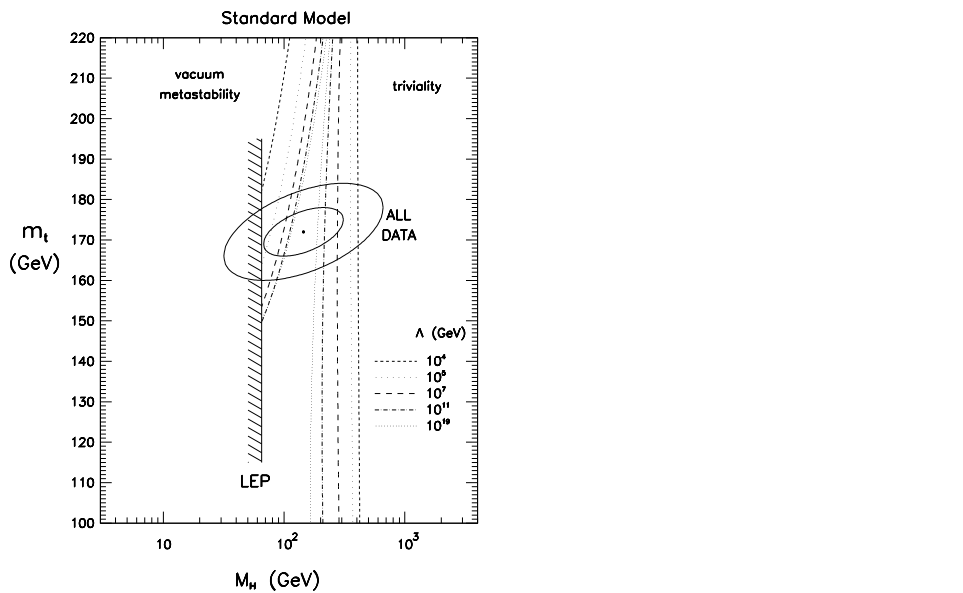,width=6.5cm}
\caption[]{
The $1$- and $2-\sigma$ predictions [42] for $M_H, m_t$ based on
precision electroweak data, confronted with the LEP lower limit and the
ranges predicted if the Standard Model is assumed to remain valid up to
different cutoff scales $\Lambda$.
}\end{figure}

\vfill\eject
{\small \begin{description}
\item{[1]} 
For a review and references, see J. Ellis and M. Karliner, CERN-TH/95-334. 
\item{[2]} 
V.N. Gribov and L.N. Lipatov, Yad. Fiz. {\bf 15} (1972) 781 [Sov. J. Nucl. Phys. {\bf
15} (1972) 438]; \\
G. Altarelli and G. Parisi, Nucl. Phys. {\bf B125} (1977) 298. 
\item{[3]}
 K. Sasaki, Prog. of Theor. Phys. {\bf 54} (1975) 1816; \\
M.A. Ahmed and G.G. Ross, Nucl. Phys. {\bf B111} (1976) 441.  
\item{[4]}
E.B. Zijlstra and W.L. van Neerven, Nucl. Phys. {\bf B417} (1994) 61; Erratum {\bf
B426} (1994) 245;\\
R. Mertig and W.L. van Neerven, Z. Phys. {\bf C70} (1996) 637; \\
W. Vogelsang, Phys. Rev. {\bf D54} (1996) 2023, RAL-TR-96-020;\\
R. Hamberg and W.L. van Neerven, Nucl. Phys. {\bf B379} (1992) 143. 
\item{[5]}
G. Mallot, talk at this meeting;\\
SMC collaboration, D. Adams et al., in preparation. 
\item{[6]}
R.L. Heimann, Nucl. Phys. {\bf B64} (1973) 429;\\
J. Ellis and M. Karliner, Phys. Lett. {\bf B213} (1988) 73. 
\item{[7]}
R.D. Ball, S. Forte and G. Ridolfi, Phys. Lett. {\bf B378} (1996) 255.
\item{[8]}
J. Bartels, B.I. Ermolaev and M.G. Ryskin, ``Non-singlet contributions to the structure
function $g_1$ at small $x$", hep-ph/9507271.
\item{[9]}
R.D. Ball and S. Forte, Nucl. Phys. {\bf B444} (1995) 287; E-ibid. {\bf B449} (1995)
680.
\item{[10]}
J. Bl\"umlein and A. Vogt, DESY-96-050 (hep-ph/9606254).
\item{[11]}
A. Sch\"afer, talk at this meeting.
\item{[12]}
J. Bjorken, Phys. Rev. {\bf 148} (1966) 1467; {\bf D1} (1970) 1376.
\item{[13]}
J. Ellis and R.L. Jaffe, Phys. Rev. {\bf D9} (1974) 1444; {\bf D10} (1974) 1669.
\item{[14]}
J. Kodaira et al., Phys. Rev.{\bf D20} (1979) 627;\\
J. Kodaira et al., Nucl. Phys. {\bf B159} (1979) 99;\\
S.A. Larin, F.V. Tkachev and J.A.M. Vermaseren, Phys. Rev. Lett. {\bf 66} (1991) 862;\\
S.A. Larin and J.A.M. Vermaseren, Phys. Lett.{\bf B259} (1991) 345;\\
S.A. Larin, Phys. Lett.{\bf B334} (1994) 192.
\item{[15}
A.L. Kataev and V.S. Starshenko, Mod. Phys. Lett. {\bf A10} (1995) 235.
\item{[16]}
The EMC Collaboration, J. Ashman et al., Phys. Lett. {\bf B206} (1988) 364; Nucl. Phys.
{\bf 328} (1989) 1.
\item{[17]}
J. Ellis and M. Karliner, Phys. Lett. {\bf B341} (1995) 397.
\item{[18]}
J. Ellis, E. Gardi, M. Karliner and M.A. Samuel, ``Pad\'e Approximants, Borel
Transforms and Renormalons: the Bjorken Sum Rule as a Case Study", hep-ph/9509312, to
be published in Phys. Lett. B.
\item{[19]}
M.A. Samuel, G. Li and E. Steinfelds, Phys. Rev. {\bf D48} (1993) 869;\\
M.A. Samuel and G. Li, Int. J. Th. Phys. {\bf 33} (1994) 1461;\\
M.A. Samuel, G. Li and E. Steinfelds, Phys. Lett. {\bf B323} (1994) 188;\\
M.A. Samuel and G. Li, Phys. Lett. {\bf B331} (1994) 114;\\
M.A. Samuel, G. Li and E. Steinfelds, Phys. Rev. {\bf E51} (1995) 3911;\\
M.A. Samuel, J. Ellis and M. Karliner, Phys. Rev. Lett. {\bf 74} No. 22 (1995) 4380.
\item{[20]}
J. Ellis, E. Gardi, M. Karliner and M.A. Samuel, CERN-TH-96/188 (hep-ph/9607404, to
be published in Phys. Rev. D).
\item{[21]}
M. Schmelling, Rapporteur talk at the {\it International Conference on High-Energy
Physics}, Warsaw, 1996.
\item{[22]}
J. Gasser and H. Leutwyler, Phys. Rep. {\bf 87} (1982) 77.
\item{[23]}
T.H.R. Skyrme, Proc. Roy. Soc. London {\bf A260} (1961) 127;\\
E. Witten, Nucl. Phys. {\bf B160} (1979) 57;\\
E. Witten, Nucl. Phys. {\bf B223} (1983) 422; ibid., 433;\\
G. Adkins, C. Nappi and E. Witten, Nucl. Phys. {\bf B228} (1983) 433;\\
For the 3-flavour extension of the model, see: E. Guadagnini, Nucl. Phys. {\bf B236}
(1984) 35; P.O. Mazur, M.A. Nowak and M. Praszalowicz, Phys. Lett. {\bf 147B} (1984)
137.
\item{[24]}
S.J. Brodsky, J. Ellis and M. Karliner, Phys. Lett. {\bf B206} (1988) 309.
\item{[25]}
A.V. Efremov and O.V. Teryaev, Dubna report JIN-E2-88-287 (1988);\\
G. Altarelli and G. Ross, Phys. Lett. {\bf B212} (1988) 391;\\
R.D. Carlitz, J.D. Collins and A.H. Mueller, Phys. Lett. {\bf B214} (1988) 219.
\item{[26]}
G.M. Shore and G. Veneziano, Phys. Lett. {\bf B244} (1990) 75;\\
G.M. Shore and G. Veneziano, Nucl. Phys. {\bf B381} (1992) 23.
\item{[27]}
FNAL E581/704 Collaboration, D.L. Adams et al., Phys. Lett. {\bf B336} (1994) 269.
\item{[28]}
R.L. Jaffe, ``Gluon Spin in the Nucleon", MIT-CTP-2466, hep-ph/9509279.
\item{[29]}
S. Forte, talk at this meeting;\\
M. Gl\"uck et al., ``Next-to-leading order analysis of polarized and unpolarized
structure functions", hep-ph/9508347;\\
T. Gehrmann and W.J. Stirling, ``Polarized parton distributions in the nucleon",
Durham preprint DTP/95/82, hep-ph/9512406;\\
See also refs. [5] and [7].
\item{[30]}
S. Okubo, Phys.Lett. {\bf 5} (1963) 165; G. Zweig, CERN Report No. 8419/TH412 (1964)
unpublished; I. Iuzuka, Prog. Theor. Phys. Suppl. {\bf 37-38} (1966) 21; see also G.
Alexander, H.J. Lipkin and P. Scheck, Phys. Rev. Lett. {\bf 17} (1966) 412.
\item{[31]}
T.P. Cheng, Phys. Rev. {\bf D13} (1976) 2161;\\
J. Gasser, M.E. Sainio and A. Svarc, Nucl. Phys, {\bf B307} (1988) 779; M.E. Sainio,
``Update of the $\sigma$ term", Helsinki report HU-TFT-95-36, Jul. 1995, invited talk
at 6th International Symposium on Meson-Nucleon Physics and Structure of the Nucleon,
Blaubeuren, Germany, 10-14 Jul. 1995.
\item{[32]}
L.A. Ahrens et al., Phys. Rev. {\bf D35} (1987) 785.
\item{[33]}
J. Ellis, E. Gabathuler and M. Karliner, Phys. Lett. {\bf B217} (1989) 173.
\item{[34]}
R.L. Jaffe, Phys. Lett. {\bf B229} (1989) 275.
\item{[35]}
OBELIX Collaboration, A. Bertin et al., hep-ex/9607006 and references therein.
\item{[36]}
J. Ellis, M. Karliner, D.E. Kharzeev and M.G. Sapozhnikov, Phys. Lett. {\bf B353}
(1995) 319.
\item{[37]}
J. Ellis, D. Kharzeev and A. Kotzinian, ``The proton spin puzzle and $\Lambda$
polarization in deep inelastic scattering", CERN-TH-95-135, hep-ph/9506280.
\item{[38]}
S. Willocq et al. (WA59 Collaboration), Z. Phys. {\bf C53} (1992) 207.
\item{39}]
J. Ellis, R. Flores and S. Ritz, Phys. Lett. {\bf 198B} (1987) 393.
\item{[40]}
R. Mayle et al., Phys. Lett. {\bf B203} (1988) 188 and {\bf B219} (1989) 515.
\item{[41]}
C. Prescott, talk at this meeting.
\item{[42]}
J. Ellis, G.L. Fogli and E. Lisi, CERN-TH/96-216 (hep-ph/9608329, to be published in
Phys. Lett. B); see also \\
A. Blondel, Plenary talk at the International Conference on High-Energy Physics,
Warsaw, 1996, reporting the analysis of the LEP Electroweak Working Group and the SLD
Heavy Flavour Group, CERN Report No. LEPEWWG/96-02, available at the URL:
http://www.cern.ch/LEPEWWG;\\
W. De Boer, A. Dabelstein, W.. Hollik, W. Moesle and U. Schwickerath, hep-ph/9609209;\\
S. Dittmaier and D. Schildknecht, hep-ph/9609488.

\end{description}}

\end{document}